\documentclass[12pt]{iopart}

\usepackage{iopams}
\usepackage{graphicx}
\usepackage{subfigure}

\begin{document}

\title[G Kandasamy et al]{Cross-talk compensation of hyperfine control in donor qubit architectures}
\author{G Kandasamy, C J Wellard and L C L Hollenberg}
\address{Centre for Quantum Computer Technology, School of Physics, University of Melbourne, Parkville, Victoria 3010, Australia}
\ead{gajendran@physics.unimelb.edu.au}

\begin{abstract}

We theoretically investigate cross-talk in hyperfine gate control of
donor-qubit quantum computer architectures, in particular the Kane proposal. By
numerically solving the Poisson and Schr\"{o}dinger equations for the
gated donor system, we calculate the change in
hyperfine coupling and thus the error in spin-rotation for the donor
nuclear-electron spin system, as the gate-donor distance is
varied. We thus determine the effect of cross-talk -- the
inadvertent effect on non-target neighbouring qubits -- which
occurs due to closeness of the control
gates (20-30nm). The use of compensation protocols is
investigated, whereby the extent of crosstalk is limited by the
application of compensation bias to a series of gates. In light of these factors the architectural
implications are then considered.

\end{abstract}

\pacs{03.67.Lx, 85.35.-p}
\submitto{\NT}
%\maketitle

\section{Introduction}

There is a growing interest in solid state proposals for quantum
computation, owing to their promise of a scalable technology given
the connection with existing nano-fabrication techniques. An
important scheme in silicon was proposed by Kane \cite{Kane1},
which although difficult to realise, has a number of advantages,
including long decoherence times of both $^{31}$P nuclear and
electron spins \cite{Tyrishkin}. Dopant based proposals, in
general propose qubits to be encoded via the spin state of the
donor-nucleus \cite{Kane1} or electron \cite{Vrijen}  or
donor-charge state \cite{Hollenberg3} of a phosphorus atom in a silicon
substrate. Single qubit control in the Kane case is achieved by the application of gate potentials (derived from surface electrodes), which shift the hyperfine
interaction of a given target qubit in or out of resonance with a global transverse AC magnetic field. Two-qubit operations are performed by
drawing together the two donor-electron wave functions, such that the
exchange interaction is modified \cite{Fang,Kettle1,Wellard1}.
This requires the donors to be separated at distances of 20-30nm.

The standard paradigm for quantum computing requires the ability
to individually control single qubits in a register of many
qubits, as well as to implement controlled inter-qubit
interactions. The proximity of neighbouring qubits results in poor localisation of gate potential, giving rise to a cross-talk effect on non-target qubits. This cross-talk in quantum control
renders selective addressability for single qubit control
problematic. A similar problem arises in other dopant based
proposals \cite{Vrijen, Skinner, Hill}, superconducting qubits and in quantum
dots\cite{Tanamoto, Snider}. Analogous to this control problem,
measurement cross-talk in such architectures may also be a problem
(similar to the SQUID equivalent \cite{McDermott}). Although both cross-talk and decoherence processes cause errors in qubit operations, we wish to distinguish the two. Decoherence is a fundamentally non-unitary process caused by interaction with the environment, whereas cross-talk leads to unwanted unitary evolution of non-target qubits during an operation. Given the stringent
requirements of control precision that fault-tolerant quantum
computation imposes at the 10$^{-4}$ level\cite{Preskill},
determination and elimination of cross-talk is a key issue for
scalable quantum computation.

This paper is organised as follows: In Sec.~2 the physical
architecture and considerations for modelling is described. In
Sec.~3 the calculation of donor-electron wave functions and
hyperfine coupling is presented with an emphasis to illustrate the
cross-talk problem and the fidelity of single qubit addressing. In Sec.~4 the gate compensation to improve
selectivity is described and considers alternative configuration of.

\section{Modelling the  Kane architecture}

\begin{figure}[h!]
\hbox{\hspace{0mm} \vspace{1mm} \parindent=0mm \vtop{ \vbox{\hsize=20mm \vspace{0mm} \hfil $a)$}}}
\centerline{\includegraphics[height=2.5in]{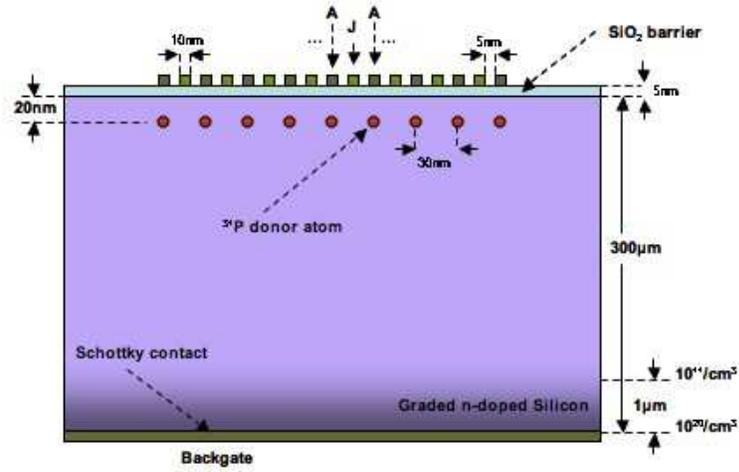}}
\hbox{\hspace{0mm} \vspace{1mm} \parindent=0mm \vtop{ \vbox{\hsize=20mm \vspace{0mm} \hfil $b)$}}}
\centerline{\includegraphics[height=1.5in]{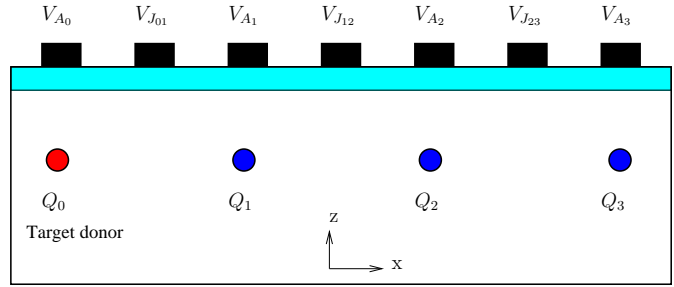}}
\caption{a) A Kane architecture based model, b) schematic configuration of gates and donors.}
\label{kane_arch}
\end{figure}

A schematic of the system we consider is shown in
Fig.~\ref{kane_arch}. This architecture consists of $^{31}$P (spin
$I=\frac{1}{2}$) atoms doped in purified $^{28}$Si ($I=0$). The
dopants are separated by 30nm and embedded at a depth of 20nm
below the 5nm oxide layer. A series of surface gates referred to
as $A$ (aligned above the donor) and $J$ (in between the donors)
are placed on a 5nm thick oxide layer. The gates are 10nm wide
and separated by 5nm. A backgate is connected to the substrate by
a graded n-doped layer such that it forms an ohmic contact with0
the substrate.

Each $^{31}$P atom forms four covalent bonds with neighbouring
$^{28}$Si atoms, with the fifth valence electron loosely bound
to the P$^+$ ion, with a Bohr radius of $a \approx$ 2nm. Operations to be performed on the system are governed by the effective spin Hamiltonian for the 2-donor system, given by:

\begin{equation}
H^{\rm spin}= H_{Z} + H_{A} + H_J +H_{\rm ac},
\label{hamiltonian}
\end{equation}

\noindent where $H_{Z}$ is the Zeeman term due to a constant magnetic field applied to the entire system
\begin{equation}
H_{Z} = \mu_BB({\bf \sigma}^{1e}_z + {\bf \sigma}^{2e}_z ) -  g_n\mu_nB({\bf \sigma}^{1n}_z +{\bf \sigma}^{2n}_z ),
\end{equation}

\noindent $H_{A}$ is the contribution of the hyperfine interaction that couples nuclear and electron spins
\begin{equation}
H_{A} =  A_1{\bf \sigma}^{1e}\cdot{\bf \sigma}^{1n} + A_2{\bf \sigma}^{2e}\cdot{\bf \sigma}^{2n},
\end{equation}

\noindent $H_J$ is the contribution of the exchange interaction coupling neighbouring electrons 
\begin{equation}
	 H_J = J_{12}{\bf \sigma}^{1e}\cdot{\bf \sigma}^{2e},
\end{equation}

\noindent and $H_{\rm ac}$ is the contribution of the transverse  rotating magnetic field to the Hamiltonian
\begin{eqnarray} 
H_{\rm ac}= -g_n\mu_n&B_{\rm ac}[(\sigma_x^{1n} +\sigma_x^{2n}) \cos(wt) +(\sigma_y^{1n} +\sigma_y^{2n}) \sin(wt)] \\ \nonumber
&+ \mu_BB_{\rm ac}[(\sigma_x^{1e}+\sigma_x^{2e}) \cos(wt) +(\sigma_y^{1e} +\sigma_y^{2e}) \sin(wt)].
\end{eqnarray}

\noindent In the above expressions ${\mu_B}$ is the Bohr magneton, ${\mu_n}$ is the nuclear magneton, ${g_n}$ is the Lande factor for $^31$P, $B$ is the magnitude of the constant magnetic field, $B_{\rm ac}$ is the magnitude of the global transverse AC magnetic field, $w$ the frequency of the transverse field and $\sigma_x$,$\sigma_y$,$\sigma_z$ the Pauli matrices operating on electron and nuclear spins. $A$ and $J$ stand for hyperfine and exchange couplings respectively. For a detailed description of the Kane architecture refer to \cite{Goan, Hollenberg2}.\\

The coupling parameters $A$ and $J$ are controlled via A and J gates, in order to perform qubit operations. The problem of cross talk can be summarised as follows; rather than being dependent on a single control gate, the Hamiltonian parameters $A,J$ are functions of biases applied to all the control gates in the vicinity:
\begin{equation}
A_0= A_0({\bf V}_{\rm bias}).
\end{equation}
\noindent where ${\bf V}_{\rm bias}=\{V_{A_0}, V_{J_{01}}, V_{A_1},...\}$ is the set of voltage biases applied on the gates (Fig.~\ref{kane_arch} b)). This dependence  and the inherent crosstalk has not received investigative attention so far.

\section{Gate control of donor-electron wave functions}
 The couplings $A$ and $J$ depend on the donor electron wave function, which is determined by the solution of the electronic wave function Hamiltonian. There have been a number of approaches to the calculation
of the gate induced shift in the hyperfine coupling strength at various
levels of sophistication of both electron wave functions and description of control field, including: perturbation theory with
envelope hydrogenic wave function for circular and strip
electrodes \cite{Larionov}, envelope hydrogenic basis
diagonalisation using numerically determined nano-gate
potentials\cite{Wellard2, Kettle2}, variational effective mass
theory in uniform fields\cite{Friesen}, tight-binding calculations
with uniform field \cite{Martins}, self-consistent calculation of
3D Poisson and DFT taking into account the six valleys of Si band
structure \cite{Lu} and numerical expansion in a basis of silicon Bloch functions in a uniform
field\cite{Wellard3}. Despite differences in detail, there is good
agreement between these various methods on the hyperfine response
to external electric fields. A hydrogenic approach has proven to be sufficient for the calculation of the hyperfine shift, whereas the details of the potential field generated by the control gate are vital to the description of the cross-talk problem.  Therefore, in this first determination of the
extent of, and possible correction procedures for cross-talk in
the quantum control problem, we adopt a hydrogenic envelope approximation for the donor electron wave function, but determine the electrostatic fields numerically to get a reliable description.

In this analysis we ignore the effects of the image charges of donor nuclei and electrons in the silicon-oxide barrier, as well as the effect of the magnetic fields on the electron wave functions. Both of these issues are discussed in \cite{Calderon}, and will alter the energies of the bound-states of the inerface well that is formed beneath the control gate. This will change the voltage at which it becomes energetically favourable to occupy these surface states for all donors, and will not produce a relative energy shift between the donors.  We also ignore any inter-donor coupling, to include this would complicate the calculation considerably as it would be necessary to calculate the ground-state of a many electron system. We expect that inter-donor coupling will only play a role when the electron wavefunctions have been significantly deformed, and we are not interested in any regime in wich more than one donor electron wavefunction is distorted.

 The ground-state single-donor wave functions are calculated by expanding in the basis of scaled hydrogen-like orbitals. The single-donor electronic Hamiltonian in the presence of an external potential generated  by a bias profile ${\bf V}_{\rm bias}$ applied to a surface gates, is given by

\begin{equation}
H^{\rm elect}=-\frac{\hbar}{2m}\nabla^2 - \frac{e^2}{4\pi\epsilon_0\epsilon_{si}r} + U(\vec{r},{\bf V}_{\rm bias}) + U_{\rm barrier}(\vec{r}).
\end{equation}

\noindent In the basis of scaled Hydrogen-like orbitals we compute the matrix elements
\begin{eqnarray}
\langle n,l,m \vert H^{\rm elect} \vert n',l',m' \rangle =\int \Psi^*_{n,l,m}({\vec r}) [ U({\vec r}, {\bf V}_{\rm bias})& + U_{\rm barrier}(\vec{r})] \Psi_{n',l',m'}({\vec r}) d{\vec r}  + \nonumber \\ & +\delta_{n,n'}\delta_{l,l'}\delta_{m,m'}\frac{E_n}{n^2}~,
\end{eqnarray}

\noindent where $n,l,m$ are the electronic, orbital and magnetic quantum numbers and  $E_n = E_1/n^2$ is the energy of the orbital. In the scaled donor units, $a_B=2{\rm nm}$ and $E_1=-45{\rm meV}$. The potential $U(\vec{r},{\bf V}_{\rm bias})$ is composed of the externally applied electrostatic potential, and $U_{\rm barrier}(\vec{r})$ is the potential barrier ($3.25\rm{eV}$) across the semiconductor-oxide interface, which arises due to the relative energies of the conduction bands of the two materials. The potentials in the device are calculated by solving the Poisson equation using ISE-TCAD \cite{tcad}. The calculations are performed for a device temperature of 1K, where convergence is obtained by bootstrapping room temperature data carefully down to required temperature. The matrix elements are calculated numerically using the Monte Carlo integration, and the Hamiltonian is diagonalised to find the donor-electron ground-state as a function of the applied gate biases.

\subsection{Fidelity of single-qubit addressing}

Single qubit operations are performed by controlling the donor nuclear-electron hyperfine coupling $A$,  via gate potentials. Both $A$, and the exchange coupling $J$, depend on the wave functions of the donor-electrons. Application of a surface gate bias deforms the donor electron wave functions, thus altering the exchange and hyperfine coupling parameters. The hyperfine coupling between the nucleus and the electron depends on the overlap of their respective wave functions. The probability density of the nucleus is treated as a delta function, so the coupling strength $A$ is given by

\begin{equation}
A({\bf V}_{\rm bias})= \frac{2\pi}{3}\mu_{\rm B}g_{\rm n}\mu_{\rm n}\mu_{\rm si} \vert\Psi(0,{\bf V}_{\rm bias})\vert^2\;,
\end{equation}

\noindent where ${\mu_{\rm si}}$ is the permeability of silicon and ${\Psi(0,{\bf V}_{\rm bias})}$ is the donor electron wave function evaluated at the nucleus \cite{Larionov}. Due to complications arising from the core electrons of the donor, it is difficult to calculate $A$ directly, rather we calculate the relative change in coupling, as a function of applied field \cite{Wellard3}:

\begin{equation}
A({\bf V}_{\rm bias}) = \frac{\vert \Psi(0,{\bf V}_{\rm bias}) \vert ^2}{\vert \Psi(0,0) \vert ^2}A(0),
\end{equation}

\noindent where $A(0)$ is determined from experimental data \cite{Fletcher}.

\begin{figure}[h!]
\hbox{\hspace{30mm} \vspace{1mm} \parindent=0mm \vtop{ \vbox{\hsize=100mm \vspace{0mm} \hfil $a) ~~~~~~~~~~~~~~~~~~~~~~~~~~~~~~~~~~~~~~~b)$}}}
\centerline{\includegraphics[height=2in]{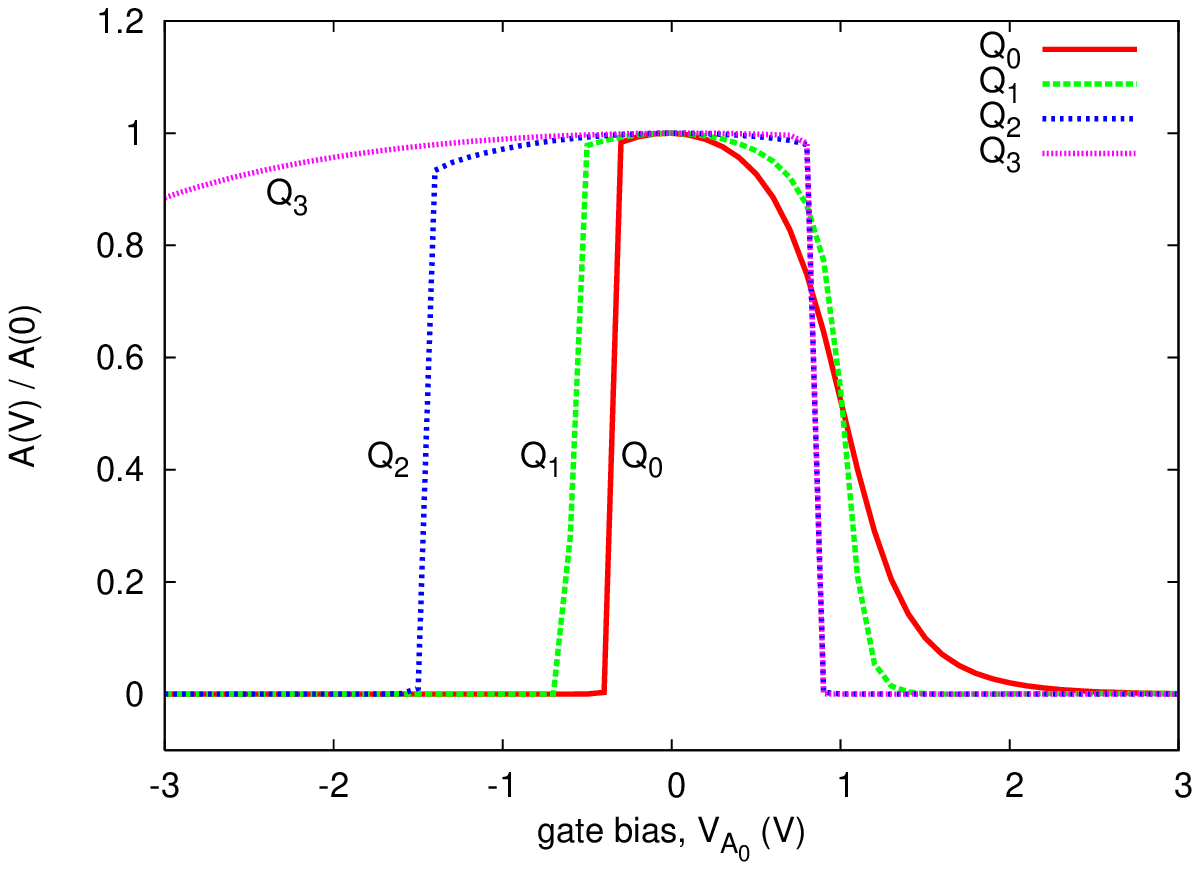}
\includegraphics[height=2in]{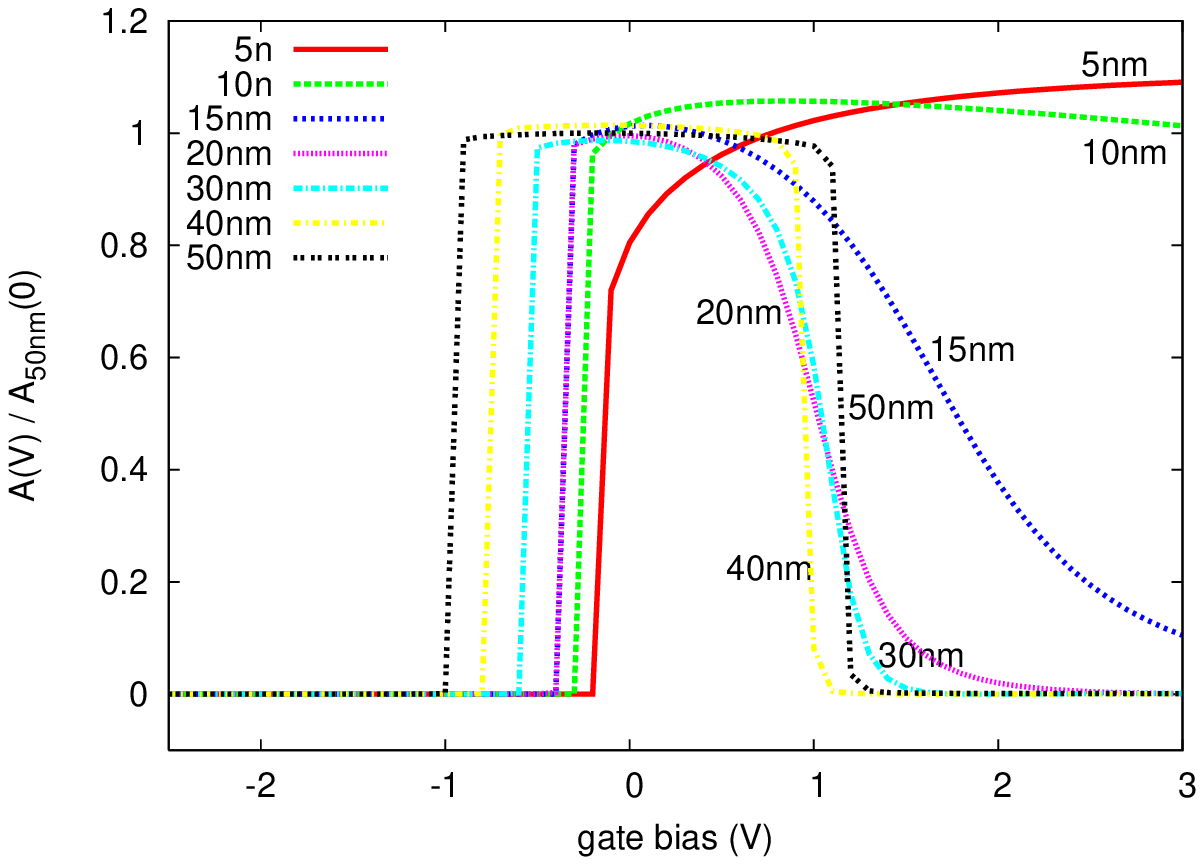}}
\caption{a)Response of the hyperfine coupling of an array of donors (as in Fig.~\ref{kane_arch} b)), b) Comparison of the Hyperfine coupling constant 'A' of target donor $Q_0$ at various depths.}
\label{basic_A}
\end{figure}

Single qubit operations are implemented by bringing the target qubit into resonance with an oscillating transverse  magnetic field. Changing $A$ changes the resonance frequency of the donor nucleus. To do this selectively requires the ability to sufficiently alter the hyperfine coupling strength of an individual donor, while keeping the other donors sufficiently far off resonance such that their spin state remains unaltered.{ As discussed, the close spacing of the qubits makes unique addressing problematic; individual hyperfine control is hindered by the gate potential cross-talk  as shown in Fig.~\ref{basic_A} a), where we have plotted the hyperfine coupling coefficient as a function of gate bias for several bits.}

The results in figure~\ref{basic_A}(b) are in good agreement with previous results \cite{Wellard2, Kettle2, Martins, Smit} and show the donor-electron response when placed extremely close to the oxide layer. The strength of the hyperfine coupling is also affected by donor depth, due to effects of both gate potential and oxide barrier. While the hyperfine coupling at donor depths $\ge$ 12nm from the oxide follow a regular pattern influenced by gate potential, very shallow depths show pronounced effects due to the oxide layer. At very shallow depths, we find that the hyperfine coupling initially increases with the applied positive bias before decreasing. This is due to the fact that when the donor is close to the oxide, the donor-electron at zero bias is pushed away from oxide, a positive bias restores $A$ to the normal value and then $A$ decreases as the wave function is pulled further towards the oxide.

\begin{figure}[h!]
\begin{center}
\hbox{\hspace{0mm} \vspace{1mm} \parindent=0mm \vtop{ \vbox{\hsize=135mm \vspace{0mm} \hfil $a) ~~~~~~~~~~~~~~~~~~~~~~~~~~~~~~~~~~~~~~~~~~~~~~~~~~~~~b)$}}}
\includegraphics[height=1.9in]{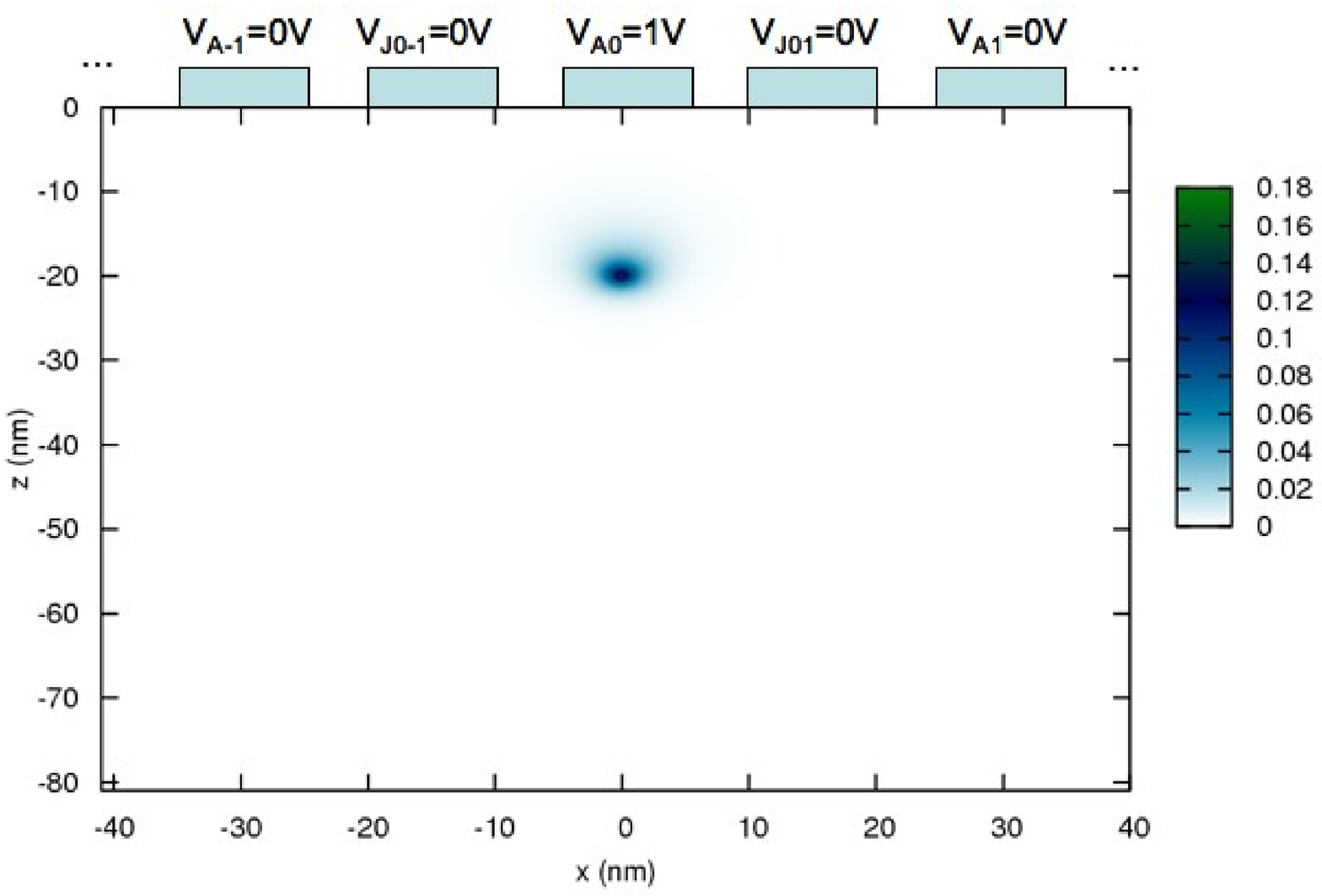}
\includegraphics[height=1.9in]{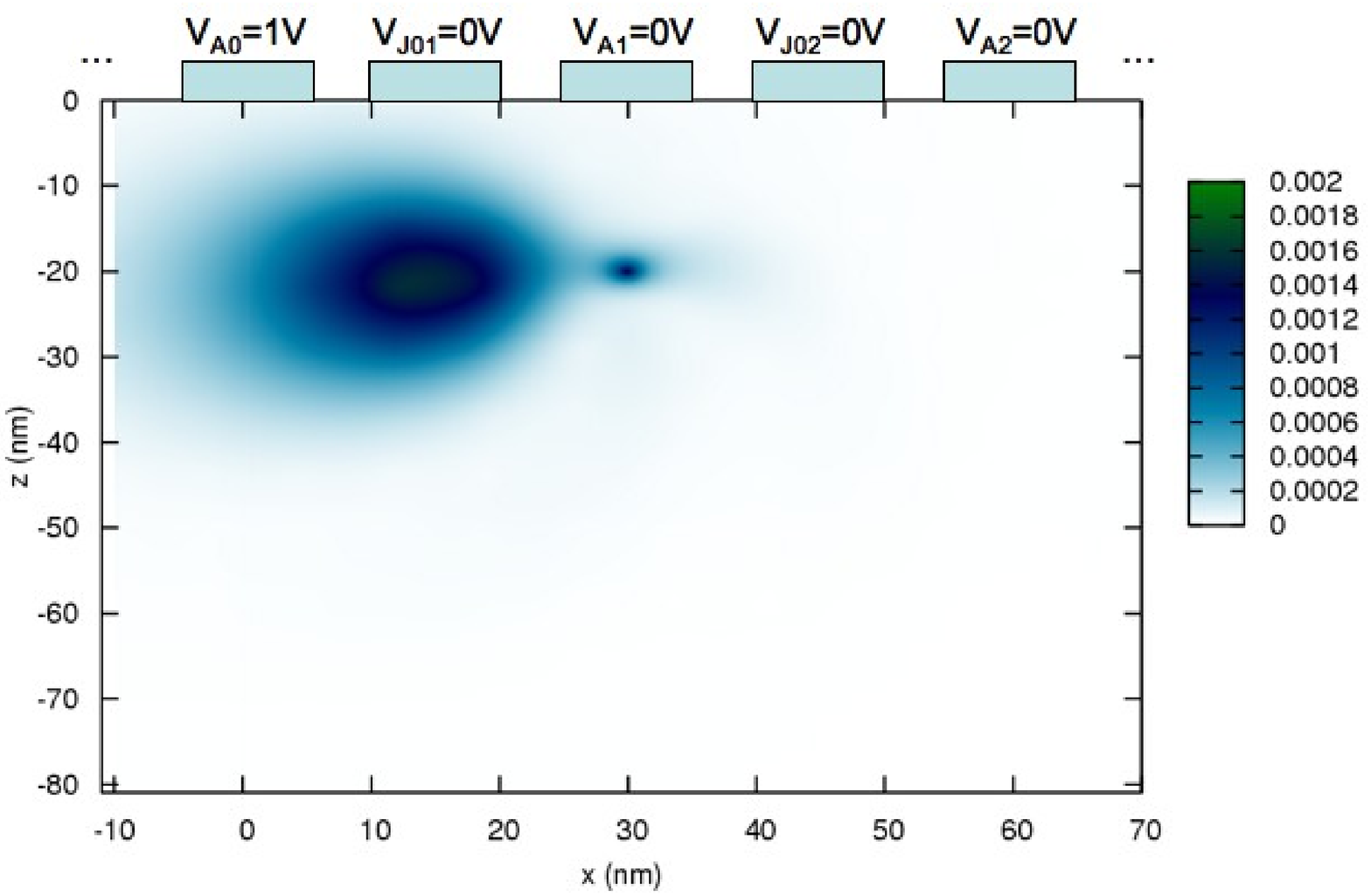}
\hbox{\hspace{0mm} \vspace{1mm} \parindent=0mm \vtop{ \vbox{\hsize=85mm \vspace{0mm} \hfil $c) ~~~~~~~~~~~~~~~~~~~~~~~~~~~~~~~~~~~~~~~~~~~~~~~~~~~~~~d)$}}}
\includegraphics[height=2in]{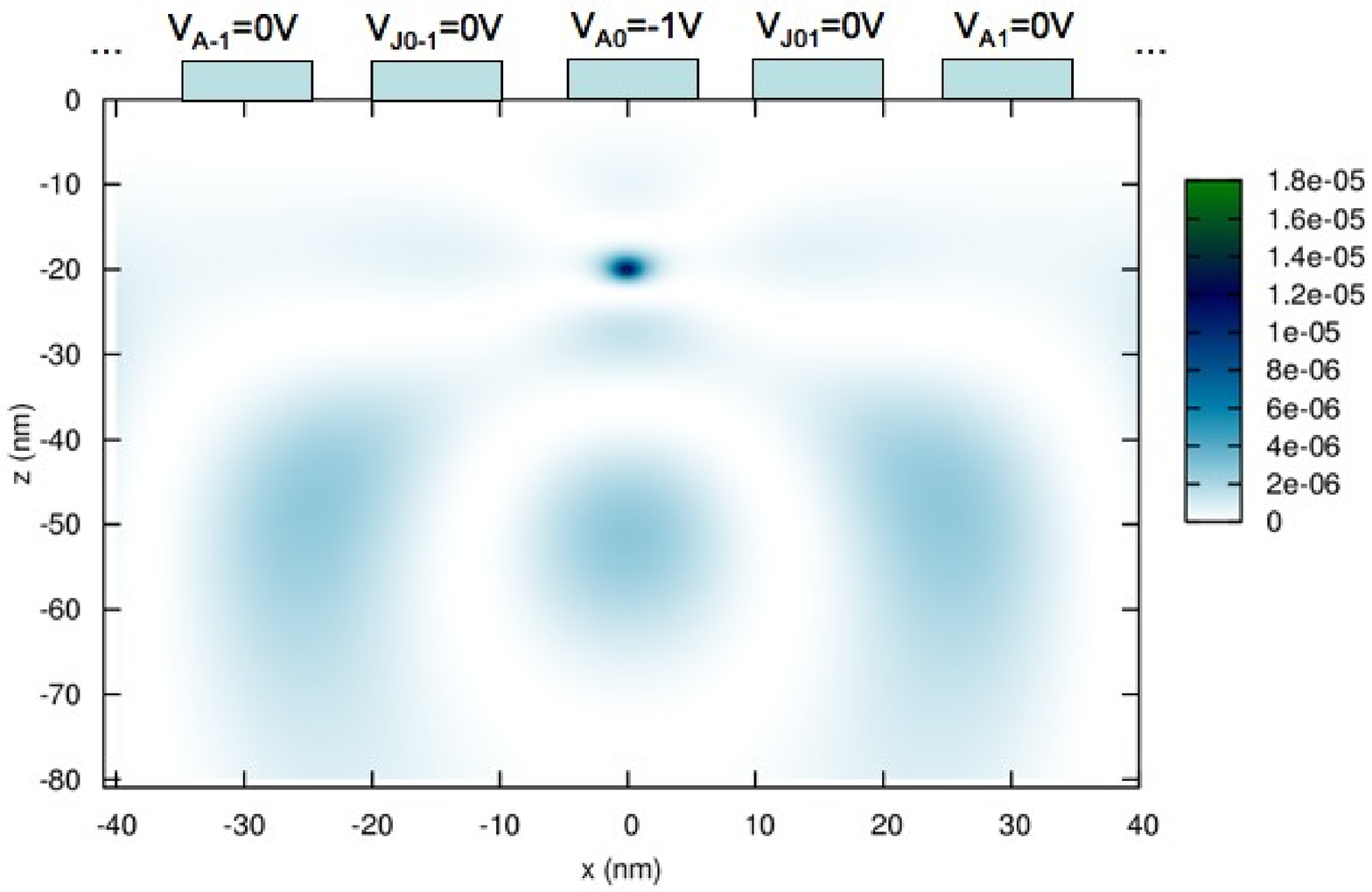}
\includegraphics[height=2in]{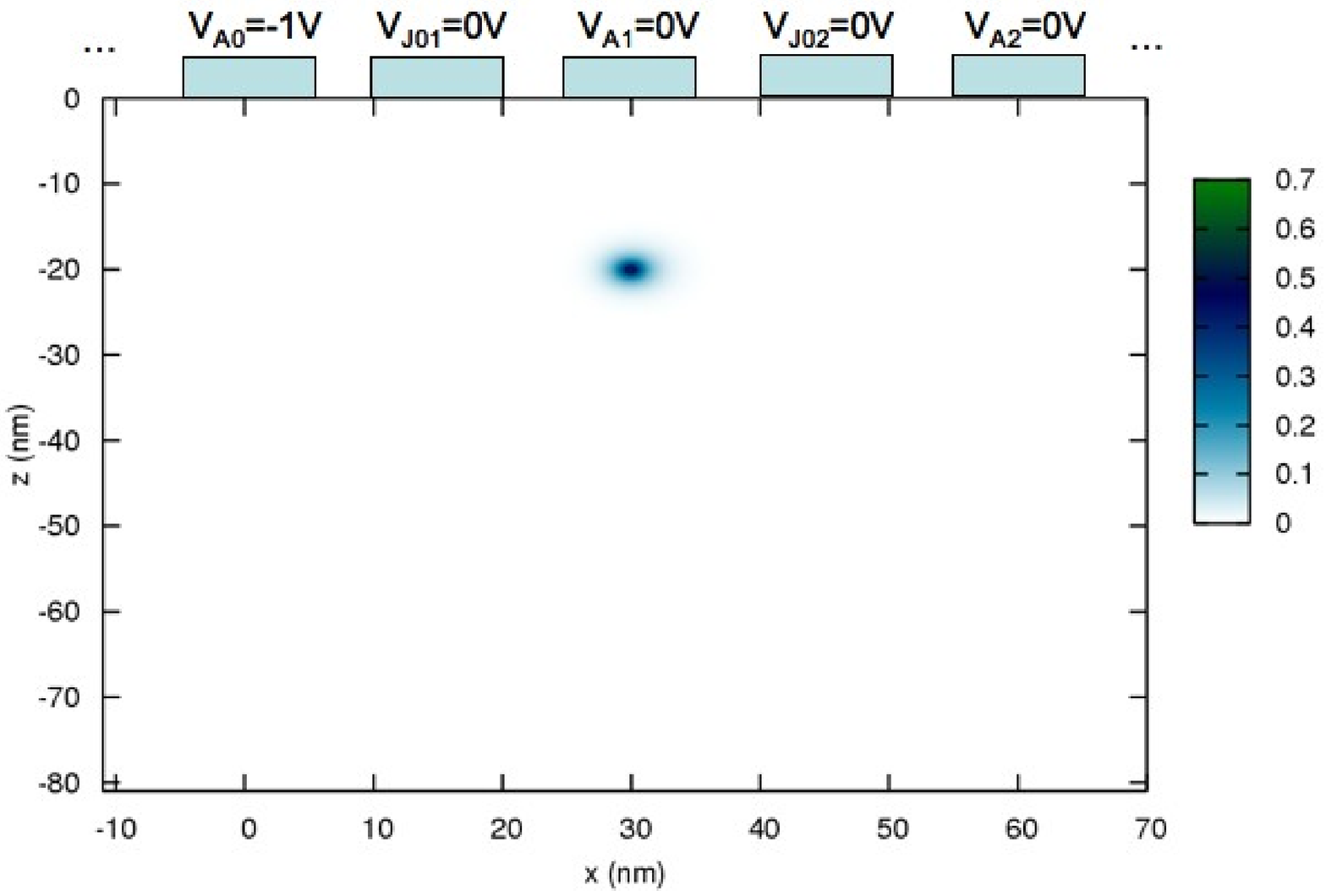}
\caption{Donor-electron probability density under perturbation when +1.0V is applied on  A$_0$ gate (all other gates are grounded) for a) target donor $Q_0$ and c) neighbouring donor $Q_1$. Negative gate voltages produces different behaviour as shown when A$_0$ gate is biased at -0.8V b) for $Q_0$ and d) for $Q_1$.}
\label{e_response}
\end{center}
\end{figure}

As the gate voltage is increased, we find that the electrons from all donors are perturbed, as shown in Fig.~\ref{e_response}. It is found that for positive applied biases the electron wave functions of neighbouring donors are perturbed at lower voltages than the target qubit, while for negative biases the behaviour is more intuitive, with the target electron perturbed at a lower bias.

\begin{figure}[h!]
\hbox{\hspace{30mm} \vspace{1mm} \parindent=0mm \vtop{ \vbox{\hsize=100mm \vspace{0mm} \hfil $a) ~~~~~~~~~~~~~~~~~~~~~~~~~~~~~~~~~~~~~~~~~~b)$}}}
\begin{center}
\includegraphics[height=2in]{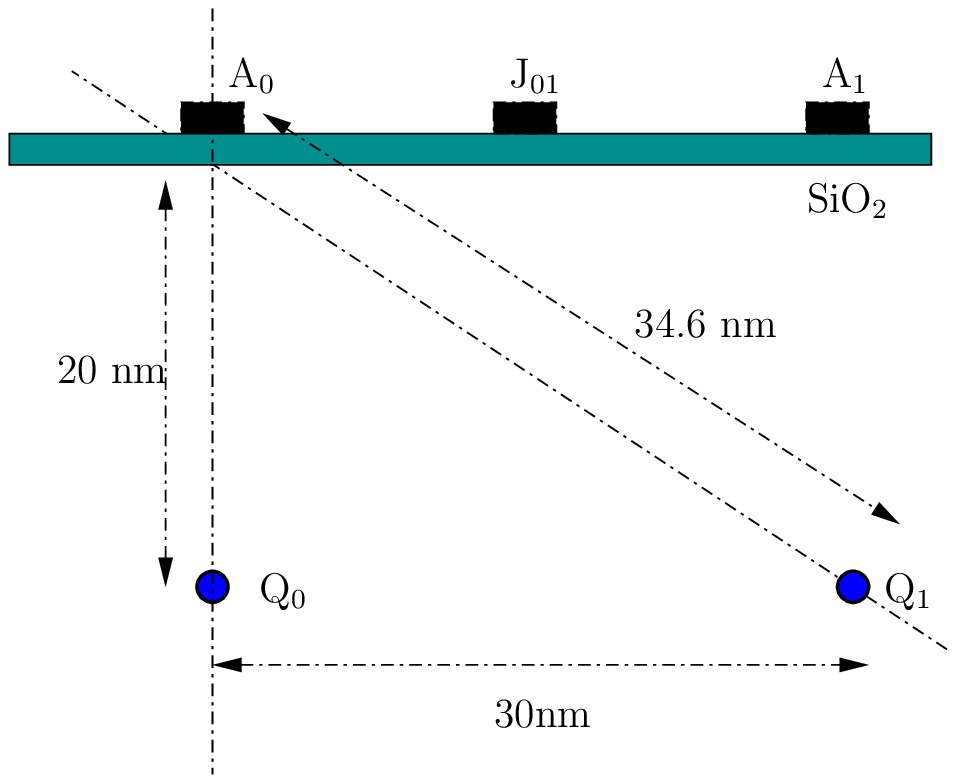}\quad\quad\quad
\includegraphics[height=2.5in]{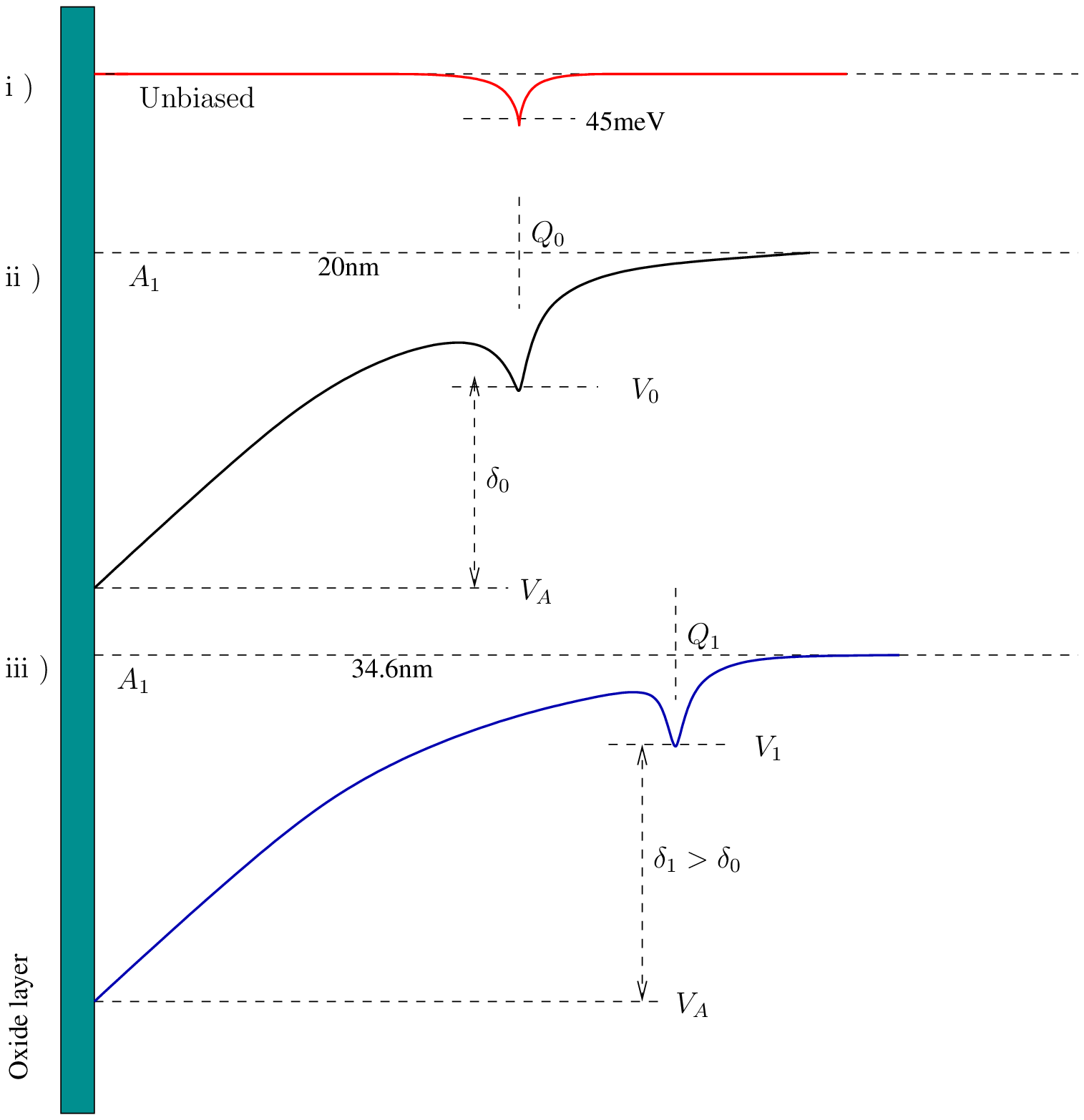}
\caption{a) The effect of gate potential increases with gate-donor distances, b) Schematic (not to scale) illustration of gate potential wells: i) potential well created by donor-electron in the absence of gate potential, ii) lowering depth of donor-electron potential well due to gate voltage, iii) lowering of depth is less at increased distance from gate, hence $\delta_1 > \delta_0$.}
\label{pot_wells}
\end{center}
\end{figure}

This can be explained as follows; the positive gate potential pulls the electron wave functions towards the oxide, while the oxide restricts the wave functions from entering it, forming a potential well at the silicon-oxide interface, underneath the gate. The depth of this potential well is the same for all donors, however donors nearer to the gate have the potential well of their ionic core lowered by the application of the field, more than for donors further from the gate. Therefore, it becomes energetically favourable for the electrons of these distant donors donors to occupy this interface well at a lower potential than is the case for donors closer to the gate (Fig.~\ref{pot_wells}).  It should be noted however, that the time scales on which these electrons tunnel into this interface well increases as the donors become further from the gate, as discussed in ref \cite{Calderon}. Therefore, it may be possible to use a kind of temporal selection whereby the gate pulses are timed such that they allow adiabatic evolution of the target donor electron into the interface well, but are too fast for the neighbouring donor electrons to tunnel. The design of such pulses with the requisite fidelity for quantum computation would be a formidable task, involving solution of he time-dependent Schr\"odinger equation in a very large basis of states. Here we restrict our attention to designing ``ground-state'' protocols, in which we assume that the pulses are applied in such a fashion that each donor electron is allowed to evolve adiabatically in its ground electronic state.

In the case of a negative applied potential, the potential wells of the donor cores are raised in energy by the gate potential, the donors close to the gate are affected more by this than those further away, and so are perturbed at a lower gate bias. However, this results in a very weakly bound donor electron state, which may be subject to significantly increased spin-orbit coupling, leading to increased dephasing rates. Additionally, such weakly bound states are more susceptible to non-reversible ionisation. We are, therefore, hesitant to propose this as a practical method for implementing gate operations.

\subsection{Effective gate errors}

Single-qubit operations are implemented by bringing the energy difference between spin states of the nucleus into resonance with a globally applied, oscillating transverse field, for a time $t$, which determines the angle of the rotation. The resonance frequency ${\nu}_{\rm res}$, of the nucleus is given by\cite{Kane1}:

\begin{equation}
h\nu_{\rm res} = g_{\rm n}\mu_{\rm n}B - \mu_{\rm B}B+\sqrt{4A^2-(g_{\rm n}\mu_{\rm n}B+\mu_{\rm B}B)^2} + 2A .
\end{equation}

At time $t$, the probability of measuring the $i^{\rm th}$ nucleus in its spin flipped state is given by  Rabi's formula \cite{Sakurai}:

\begin{equation}
P_i(t)=\frac{\gamma^2}{\gamma^2 + (\nu-\nu_{\rm res}^i)^2/4}\rm{sin}^2\left[t\sqrt{\left( \gamma^2 + \frac{\nu-\nu_{\rm res}^i}{4}\right)}\right],
\end{equation}

\noindent where $\nu$ is the frequency of the field, $\nu_{\rm res}^i$ is the resonant frequency of the nucleus and ${\gamma = 2g_n\mu_nB_{ac}/h}$. For the target qubit $Q_0$, we define the error as $\epsilon_0 = 1 - {\rm max}[P_0(t)]$, which gives the error due to under-rotation of the qubit. In the case of the non-target qubit (off-resonant with the RF field in general), we define the error as $\epsilon_{i \neq 0} = {\rm max}[P_i(t)]$, which quantifies any unwanted rotation of the qubit. In what follows we consider the error in implementing spin flip on a particular target qubit,  we assume that there is no error associated with the timing of the pulse and so set the sine function in the above expression to unity.

Studies in fault tolerant quantum computation provide a rough estimate of the allowable error in any gate operation, which is no more than one part in $10^4$ \cite{Preskill}. In the case of single qubit rotations, the error is composed both of contributions associated with the target qubit being out of resonance, as well as neighbouring qubits being insufficiently off resonance so that they undergo a partial rotation.

\begin{figure}[h!]
\begin{center}
\hbox{\hspace{40mm} \vspace{0mm} \parindent=0mm \vtop{ \vbox{\hsize=0mm \vspace{0mm} \hfil $a)~~~~~~~~~~~~~~~~~~~~~~~~~~~~~~~~~~~~~~~~~~~~~~~~~~~~~~~~~~~~~~~~~~~~~~~~~~~~~~~~~~~~~b)$}}}
\centerline{\includegraphics[height=2in]{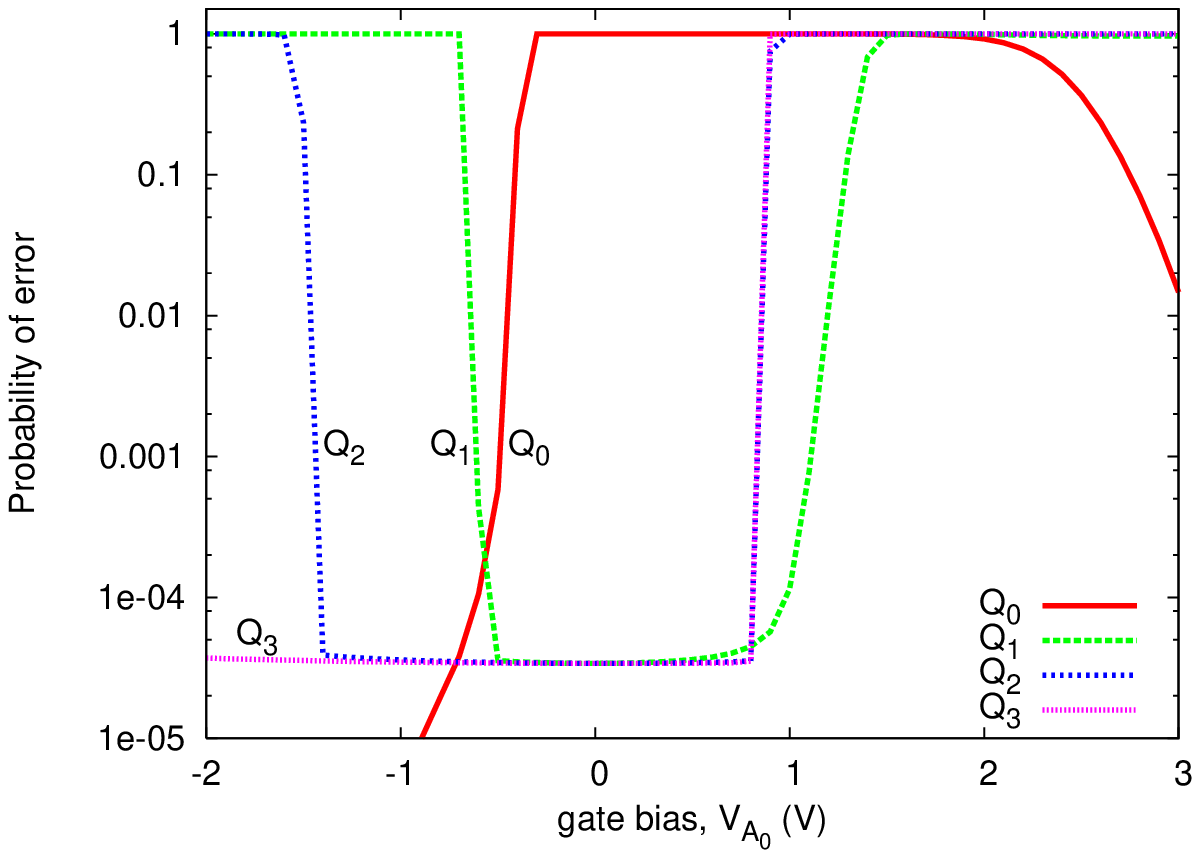}
\includegraphics[height=2in]{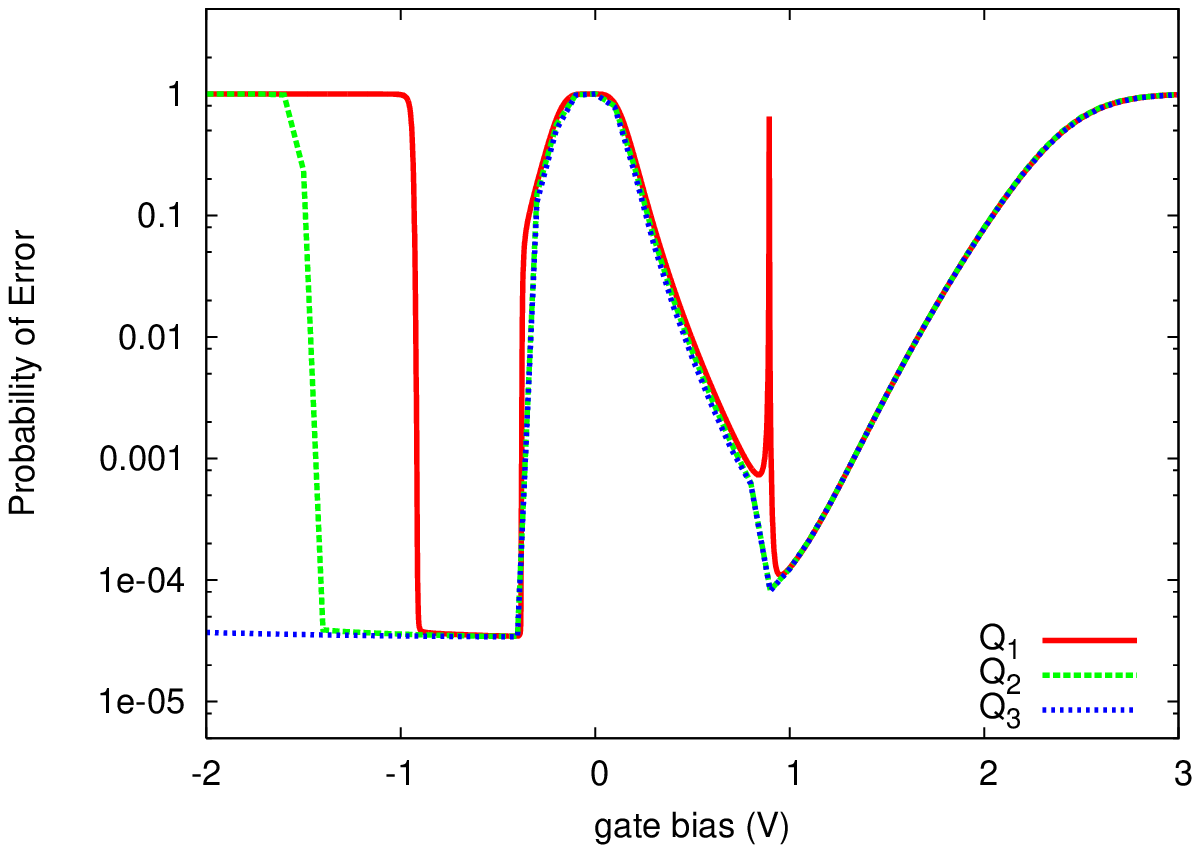}}
\caption{Evaluating performance of single qubit operation on a qubit array : a) when resonance frequency is chosen to be nuclear frequency of target when $A_0$=0, b) when resonance frequency is chosen to be the nuclear frequency of the target donor ($Q_0$ does not have any error and hence not plotted). }
\label{error_f}
\end{center}
\end{figure}

The hyperfine coupling is least sensitive to gate voltage fluctuations when $\partial{A}/\partial{V}$=0, which is the case when either  $A=A(0)$, or, with the exception of very shallow depths, for $A=0$ \cite{Kane2}, as seen in (Fig.~\ref{basic_A} b)). This suggests that the frequency of the global field should be chosen to be the nuclear resonance frequency of the target qubit in one of those cases. We set the AC field to be resonant with the $A=0$ case, in the first instance. Plotted in Fig.~\ref{error_f} a) is the gate error for each qubit as we change the target gate bias, at this frequency. We find that the required error rates are not achieved for positive gate biases.

To determine optimum frequency for the global field, for the implementation of single qubit operations, we allow the global field to be the nuclear frequency of the target qubit ($\nu=\nu_{res}$) and sweep the gate bias for the relevant range. These results are plotted in  (Fig.~\ref{error_f} b)), and show that  positive biases do not give a practical window of operation. The spike at around 1V is due to the crossover of the hyperfine coupling between target and neighbouring donor-electrons (Fig.~\ref{basic_A} a)). It is clear that some kind of compensation is required to reduce the cross-talk.

\section{Compensation for cross-talk induced gate errors}

The diffuse nature of the electric fields produced inside the device by the application of a bias voltage to any particular control gate makes selective addressability of qubits problematic. In this section we investigate a simple linear compensation scheme to address this issue. Through the application of compensation biases to neighbouring gates we minimise the electric potential in the vicinity of non-target donors whilst maintaining as much as possible in the vicinity of the target qubit. Therefore we design a bias profile such that both the potential, and x-component of the electric field, at the positions of the non-target qubits be zero. The required compensation gate profile is achieved by the following procedure. Assuming linearity of potentials in the device, the potential and it's x-gradient, at the donor sites can be described as:

\begin{equation}
{\bf \Pi} = M {\bf V}_{\rm bias}
\label{compmain}
\end{equation}

\noindent where ${\bf V}_{\rm bias}$ refer to the gate biases, and ${\bf \Pi}$ refer to potentials and their gradients at the respective qubits:
\begin{equation}
{\bf \Pi} = (U_{Q_0},U'_{Q_0},U_{Q_1},U'_{Q_1},...)^T \quad.\\ \nonumber
\end{equation}

\noindent Here
\begin{equation}
U_{Q_0} \equiv U(\vec{r}_{Q_0}; {\bf V}_{\rm bias}) = u_{A_0}(\vec{r}_{Q_0})V_{A_0} + u_{J_{01}}(\vec{r}_{Q_0})V_{J_{01}} + ... \quad , \\ \nonumber
\end{equation}

\noindent and

\begin{equation}
u_{A_0}(\vec{r}_{Q_0}) = U(\vec{r};\{V_{A_0}=1,0,0,...\}) \quad ,\nonumber\end{equation}

\noindent with $u_{A_0}(\vec{r}_{Q_0})$, the potential at qubit $Q_0$ due a bias $V_{A_0}=1V$. The matrix $M$ is constructed by solution of the Poisson equation using a commercial software package TCAD, such that Eqn.~\ref{compmain} can be written as follows

\begin{equation}
\left(\begin{array}{c}
U_{Q_0}\\
U_{Q_1}\\
U'_{Q_1}\\
\vdots\\
U_{Q_n}\\
U'_{Q_n}
\end{array} \right) = \left(\begin{array}{ccccc}
u_{A_0}(\vec{r}_{Q_0})&u_{J_{01}}(\vec{r}_{Q_0})&u_{A_1}(\vec{r}_{Q_0})&\cdots&u_{A_n}(\vec{r}_{Q_0})\\
u_{A_0}(\vec{r}_{Q_1})&u_{J_{01}}(\vec{r}_{Q_1})& & & \\
u'_{A_0}(\vec{r}_{Q_1})& & & &\vdots\\
\vdots& &\ddots& &\vdots\\
u_{A_0}(\vec{r}_{Q_n})& & &\ddots &\vdots\\
u'_{A_0}(\vec{r}_{Q_n})&\cdots &\cdots &\cdots &u'_{A_n}(\vec{r}_{Q_n})
\end{array} \right) \left(\begin{array}{c}
V_{A_0}\\
V_{J_{01}}\\
V_{A_1}\\
\vdots\\
V_{J_{(n-1)n}}\\
V_{A_n}
\end{array} \right) ,
\label{Mmatrix}
\end{equation}

\noindent where $U'$ is the derivative of potential with respect to x-direction. We require a value of ${\bf V}_{\rm bias}$, such that gives a ${\bf \Pi}$ with $U_{Q_0} =1$, and all other elements zero. The gradient of the potential at the target donor does not need compensation due to symmetry, and hence is ignored. This is obtained by matrix inversion;

\begin{equation}
{\bf V}_{\rm bias}= M^{-1}{\bf \Pi}.
\end{equation}

\begin{figure}[h!]
\hbox{\hspace{40mm} \vspace{1mm} \parindent=0mm \vtop{ \vbox{\hsize=0mm \vspace{0mm} \hfil $a)~~~~~~~~~~~~~~~~~~~~~~~~~~~~~~~~~~~~~~~~~~~~~~~~~~~~~~~~~~~~~~~~~~~~~~~~~~~~~~~~~~~b) $}}}
\centerline{\includegraphics[height=3in, angle=-90]{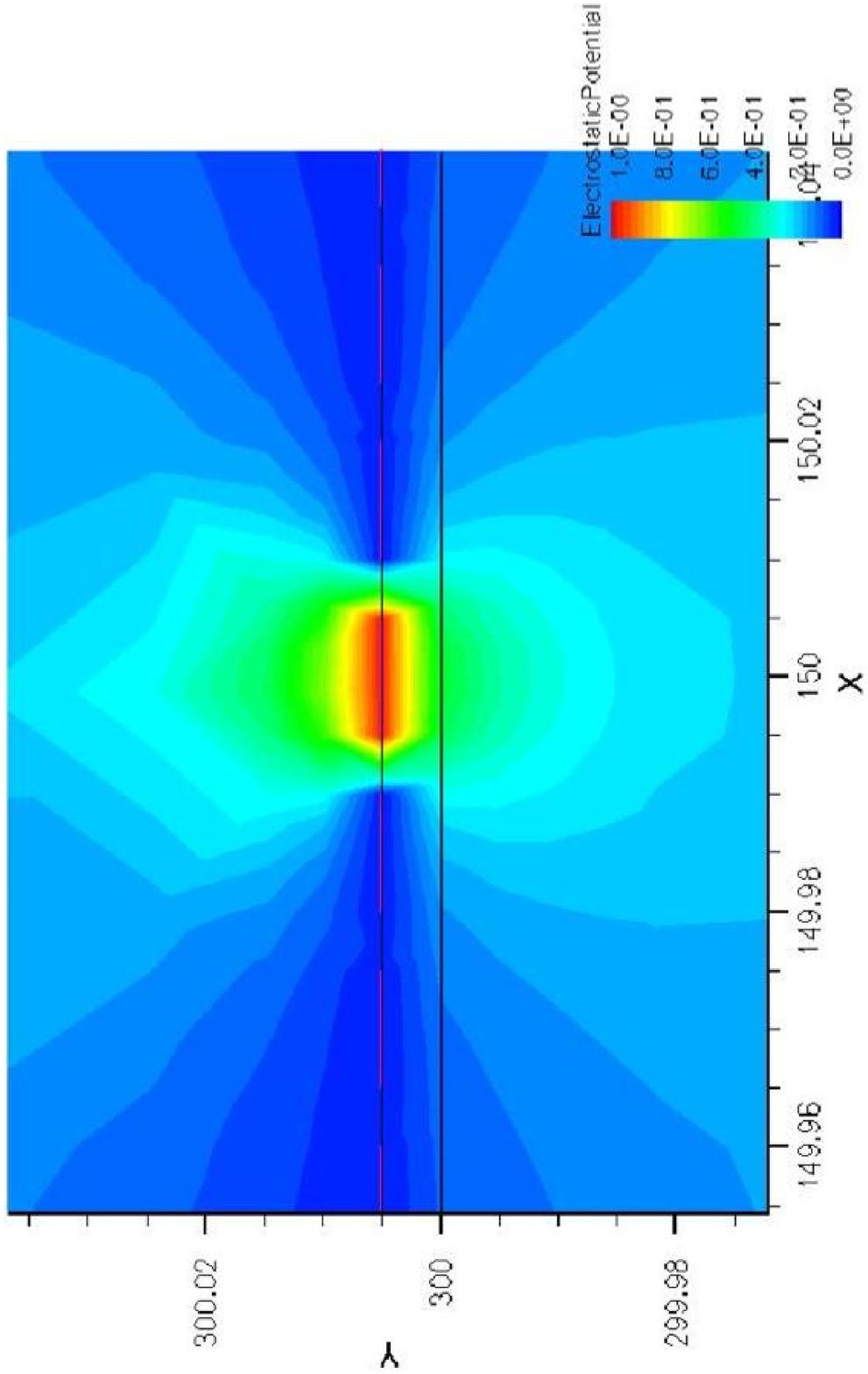}
\includegraphics[height=3in, angle=-90]{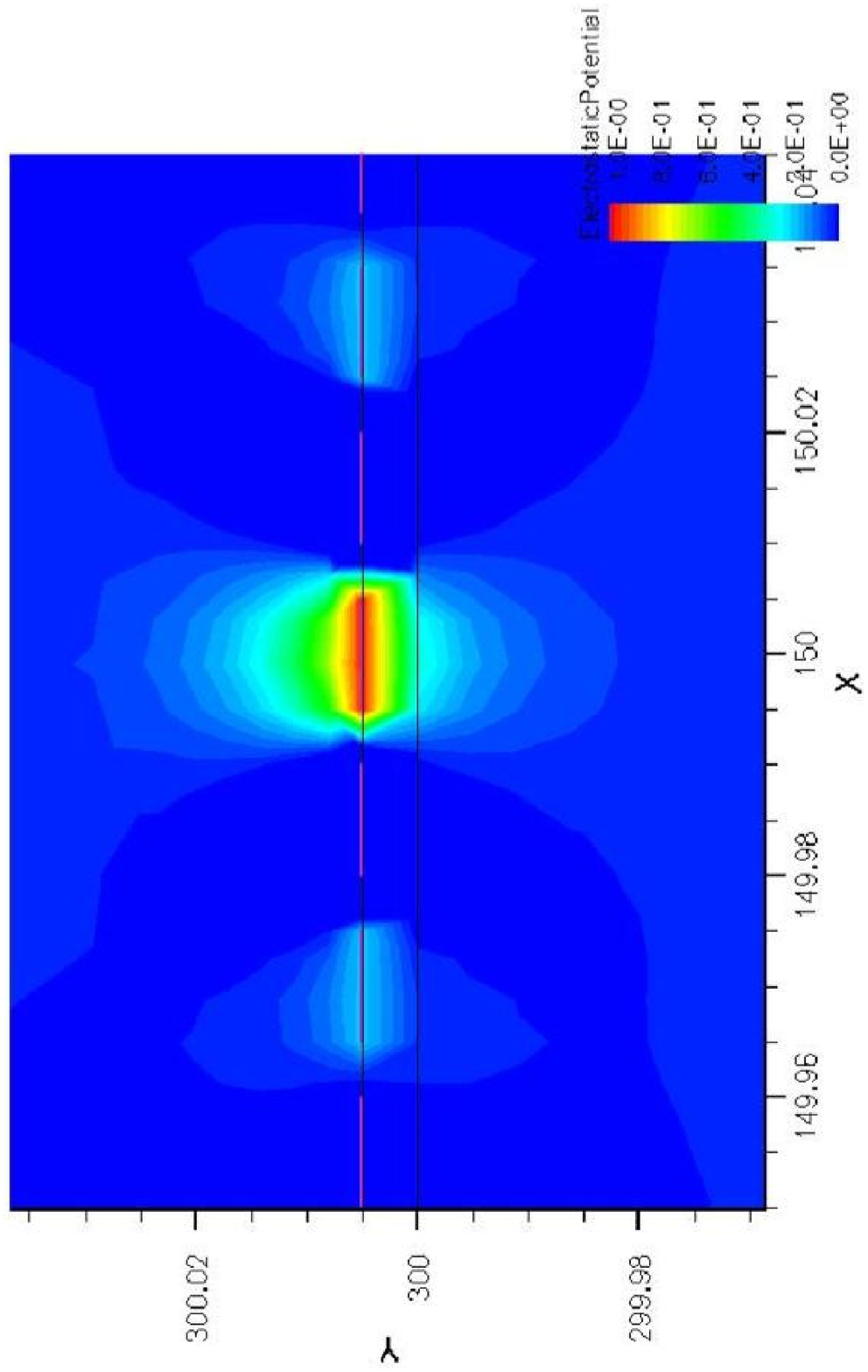}}
\caption{Potential profiles obtained from Poisson solver: a) one gate is biased while the rest are grounded b) after compensation, the potentials directed at the target is  more localised.}
\label{comp_top}
\end{figure}

\begin{figure}[h!]
\hbox{\hspace{40mm} \vspace{0mm} \parindent=0mm \vtop{ \vbox{\hsize=0mm \vspace{0mm} \hfil $a) ~~~~~~~~~~~~~~~~~~~~~~~~~~~~~~~~~~~~~~~~~~~~~~~~~~~~~~~~~~~~~~~~~~~~~~~~~b)$}}}
\centerline{\includegraphics[height=2in]{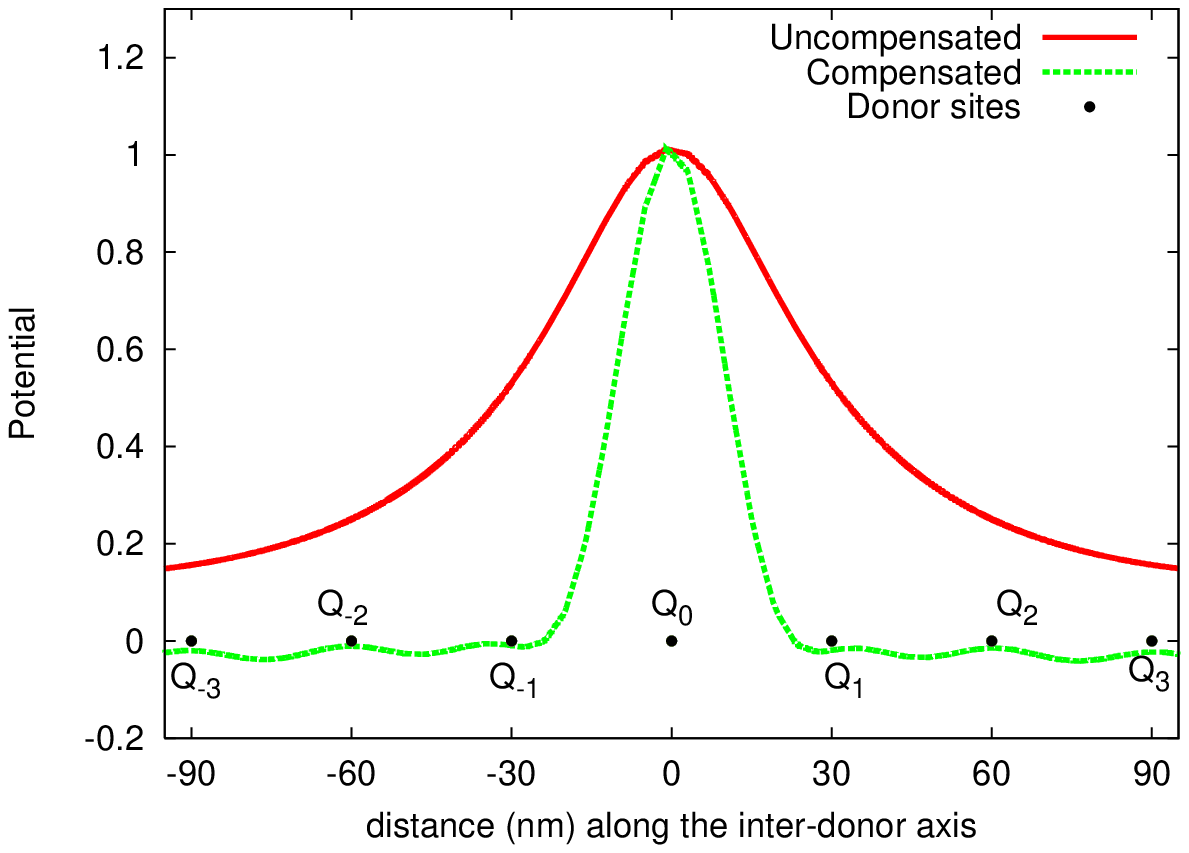}
\includegraphics[height=2in]{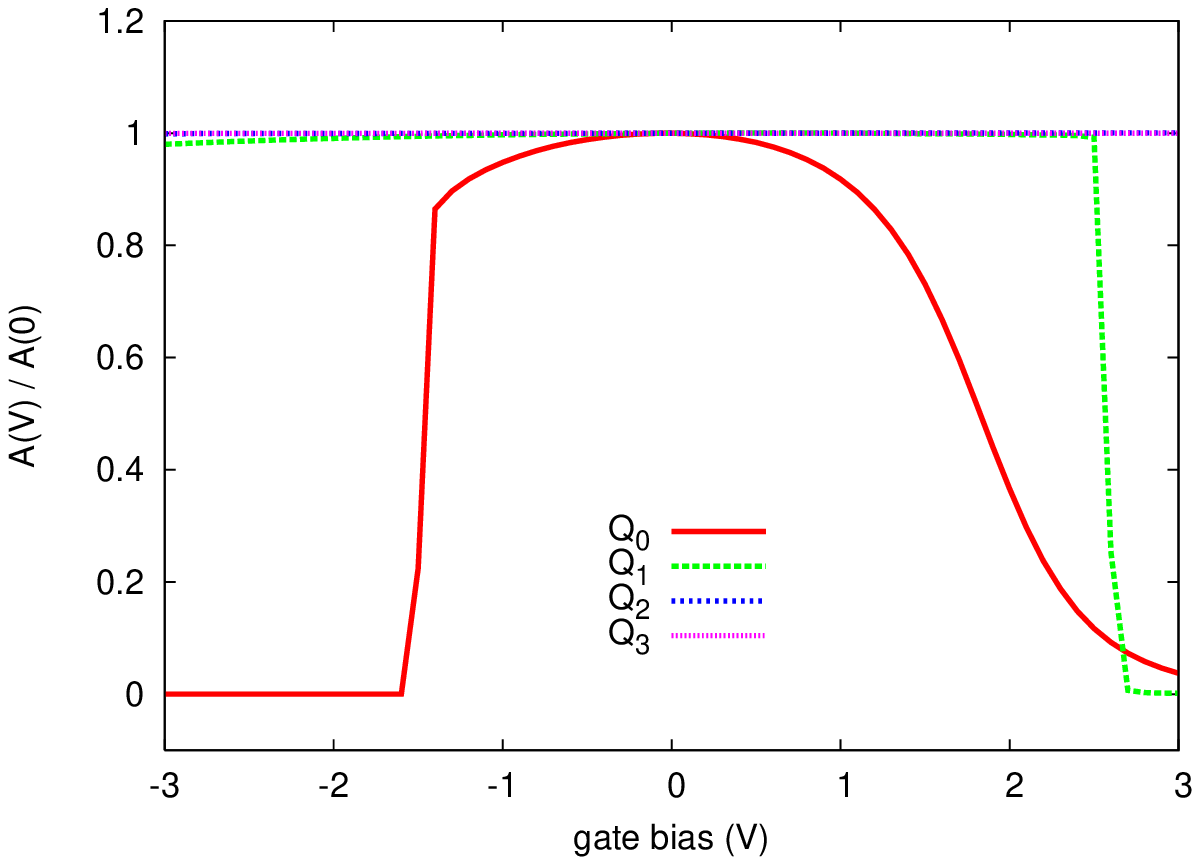}}
\caption{a) Potential profile before and after application of compensation biases to other gates, b)Hyperfine coupling of donor array after compensation with respect to $V_{A_0}$.}
\label{comp_A}
\end{figure}

Applying the above compensated bias profile produces a potential that is more localised around the target donor (Fig.~\ref{comp_top} b)). The compensation assumed linearity of the medium, which would be the case in the absence of charge carriers. However non-linearities in potentials introduced by the presence of P-dopants ($10^{11}/{\rm cm}^3$) in the semiconductor, result in slight deviation from ideal compensation at all donor-sites, as can be seen in  (Fig.~\ref{comp_A} a)). 

\begin{figure}[h!]
\hbox{\hspace{40mm} \vspace{0mm} \parindent=0mm \vtop{ \vbox{\hsize=0mm \vspace{0mm} \hfil $a) ~~~~~~~~~~~~~~~~~~~~~~~~~~~~~~~~~~~~~~~~~~~~~~~~~~~~~~~~~~~~~~~~~~~~~~~~~b)$}}}
\centerline{\includegraphics[height=2in]{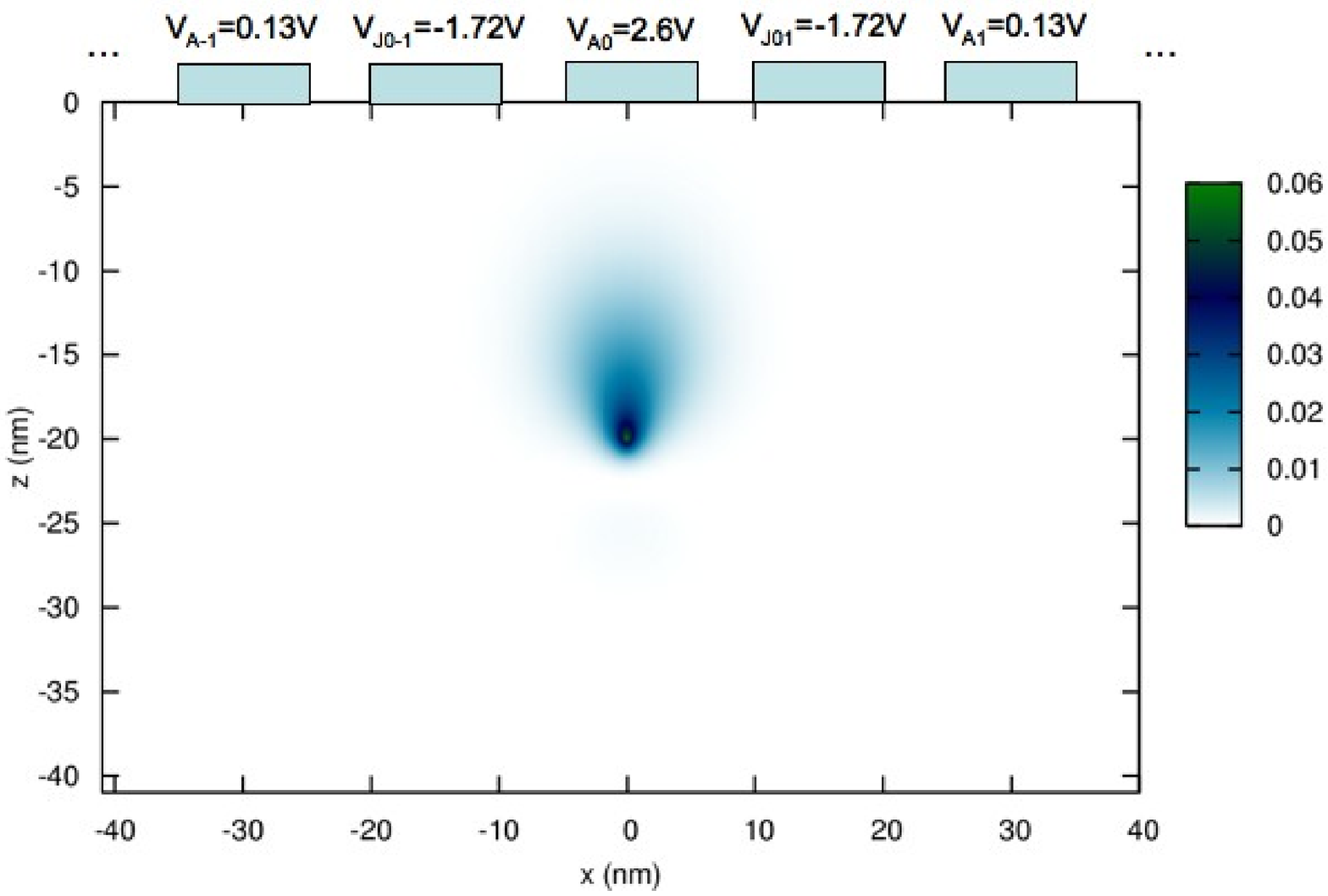}
\includegraphics[height=2in]{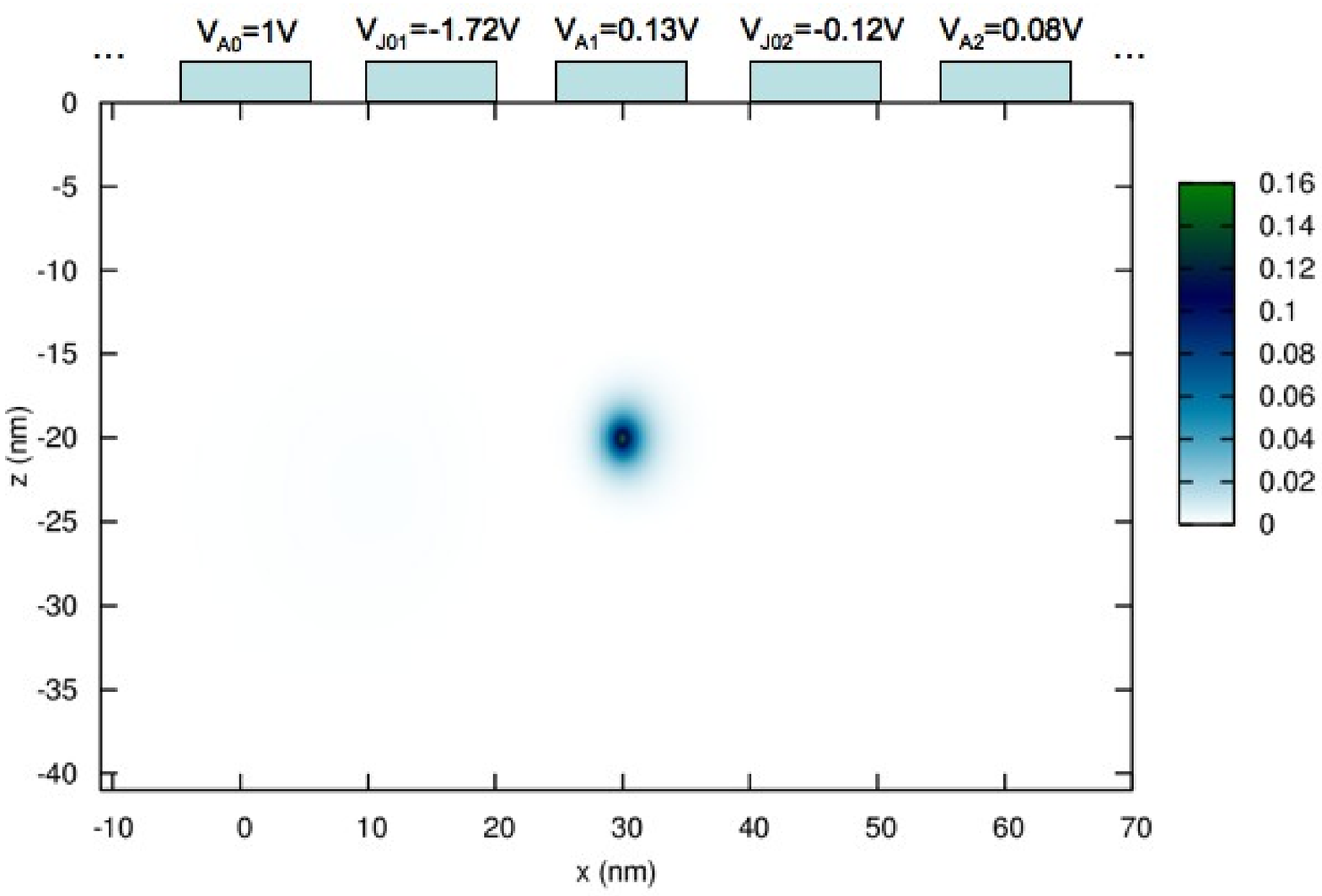}}
\caption{ Wave function response of a) $Q_0$, b) $Q_1$ after compensation.}
\label{dens_comp}
\end{figure}

The response of electron-nuclear hyperfine coupling to compensation biases is plotted in  Fig.~\ref{comp_A} b). One of the consequences of compensation is that larger voltages are required to affect the donors, with a 2V bias required to produce the same effect as 1V in the uncompensated case.  This is due to the application of counter-acting voltages on the neigbouring gates, which have the effect of reducing the overall potential within the device. In (Fig.~\ref{dens_comp}) we show the probability density for target and neighbouring donor electrons with a compensated bias of 2.6V applied to the target gate $V_{\rm A_0}$. This shows that, in contrast to the uncompensated case, the target qubit feels a greater effect than the neighbour, as required.

\begin{figure}[h!]
\hbox{\hspace{40mm} \vspace{0mm} \parindent=0mm \vtop{ \vbox{\hsize=0mm \vspace{0mm} \hfil $a) ~~~~~~~~~~~~~~~~~~~~~~~~~~~~~~~~~~~~~~~~~~~~~~~~~~~~~~~~~~~~~~~~~~~~~~~~~b)$}}}
\centerline{\includegraphics[height=2in]{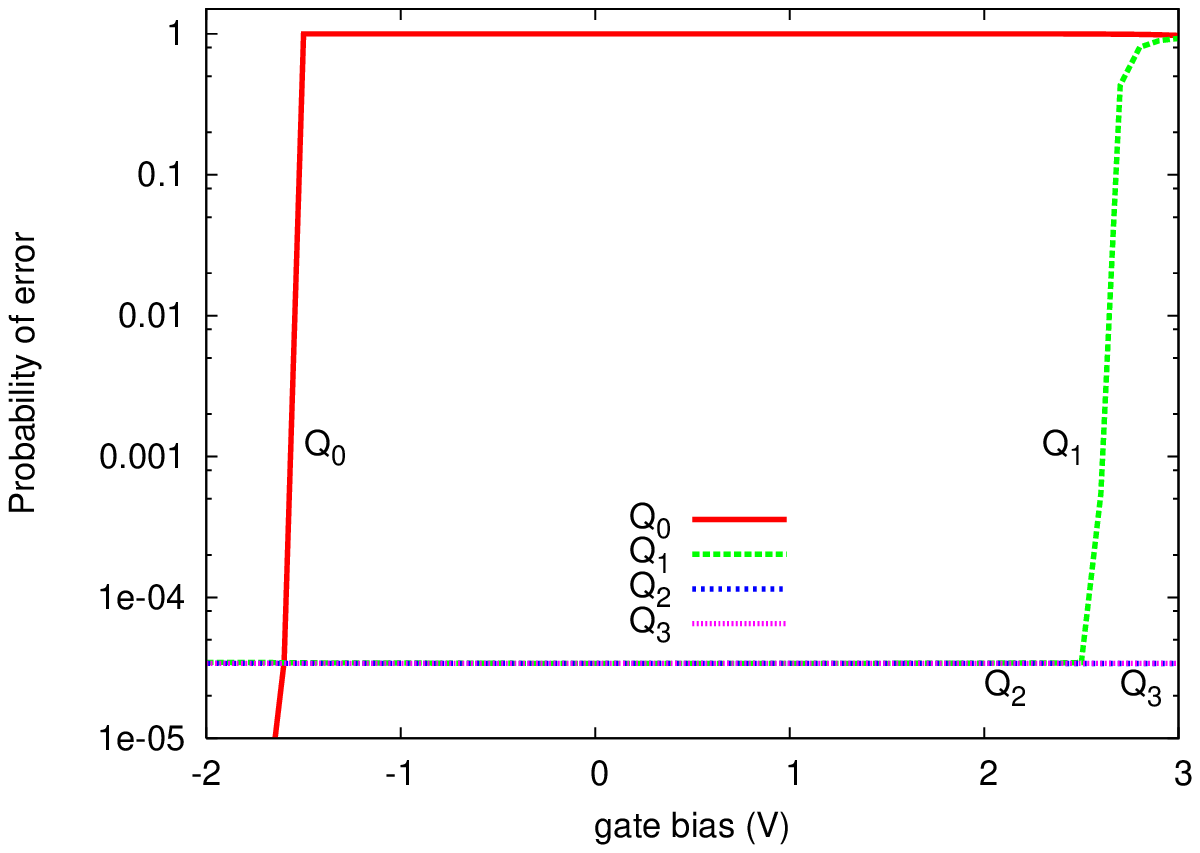}
\includegraphics[height=2in]{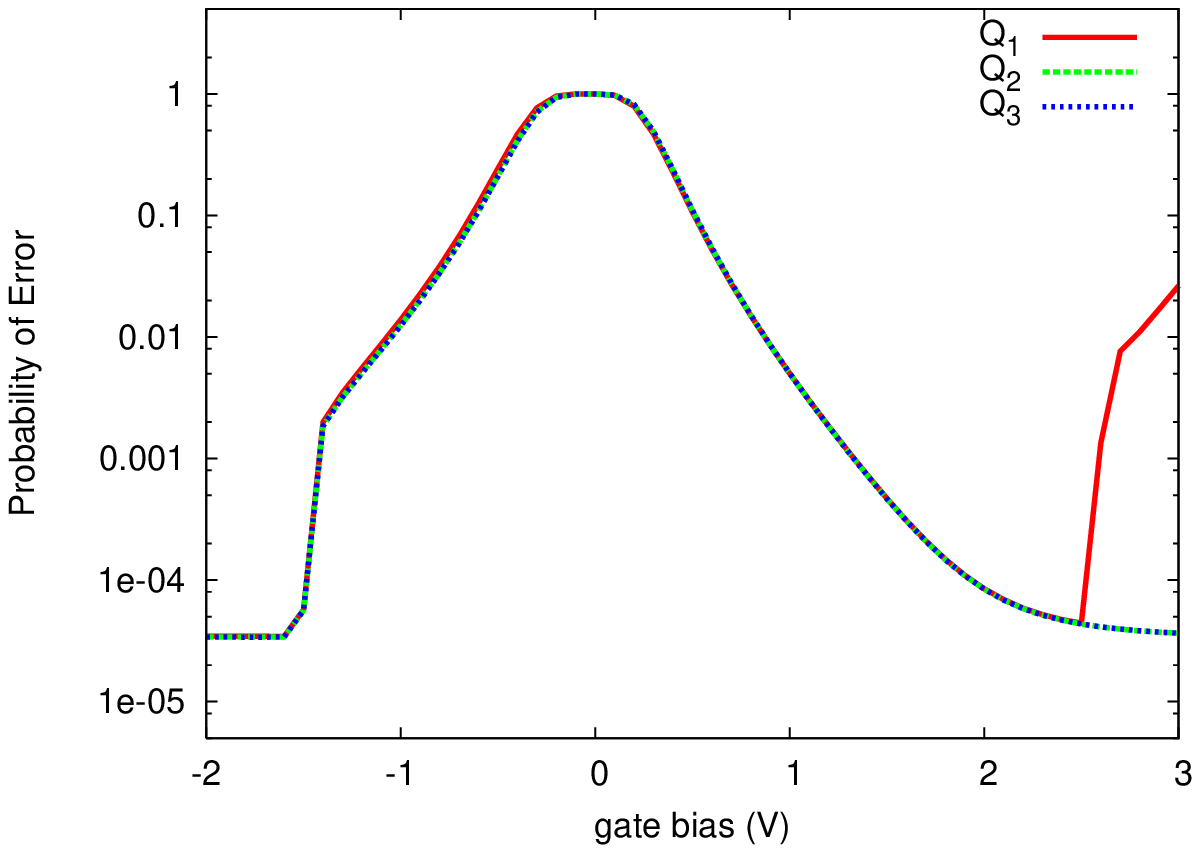}}
\caption{Error probability : When resonance frequency is chosen to be nuclear frequency of target when $A_0$=0 , b)When resonance frequency is chosen to be the nuclear frequency of the target donor ($Q_0$ does not have any error by definition in this case and hence not plotted).}
\label{error_ac}
\end{figure}

Evaluating the fidelity of a single qubit rotation on a compensated qubit array, shows a marked difference to the uncompensated case (Fig.~\ref{error_ac}). In the positive voltage regime, there exists a small window, at about 2V - 2.5V, in which the required fidelity is achieved (Fig.~\ref{error_ac} b)). However, this may still be too constrained.

\begin{figure}[h!]
\hbox{\hspace{40mm} \vspace{0mm} \parindent=0mm \vtop{ \vbox{\hsize=0mm \vspace{0mm} \hfil $a) ~~~~~~~~~~~~~~~~~~~~~~~~~~~~~~~~~~~~~~~~~~~~~~~~~~~~~~~~~~~~~~~~~~~~~~~~~~b)$}}}
\centerline{\includegraphics[height=1.8in]{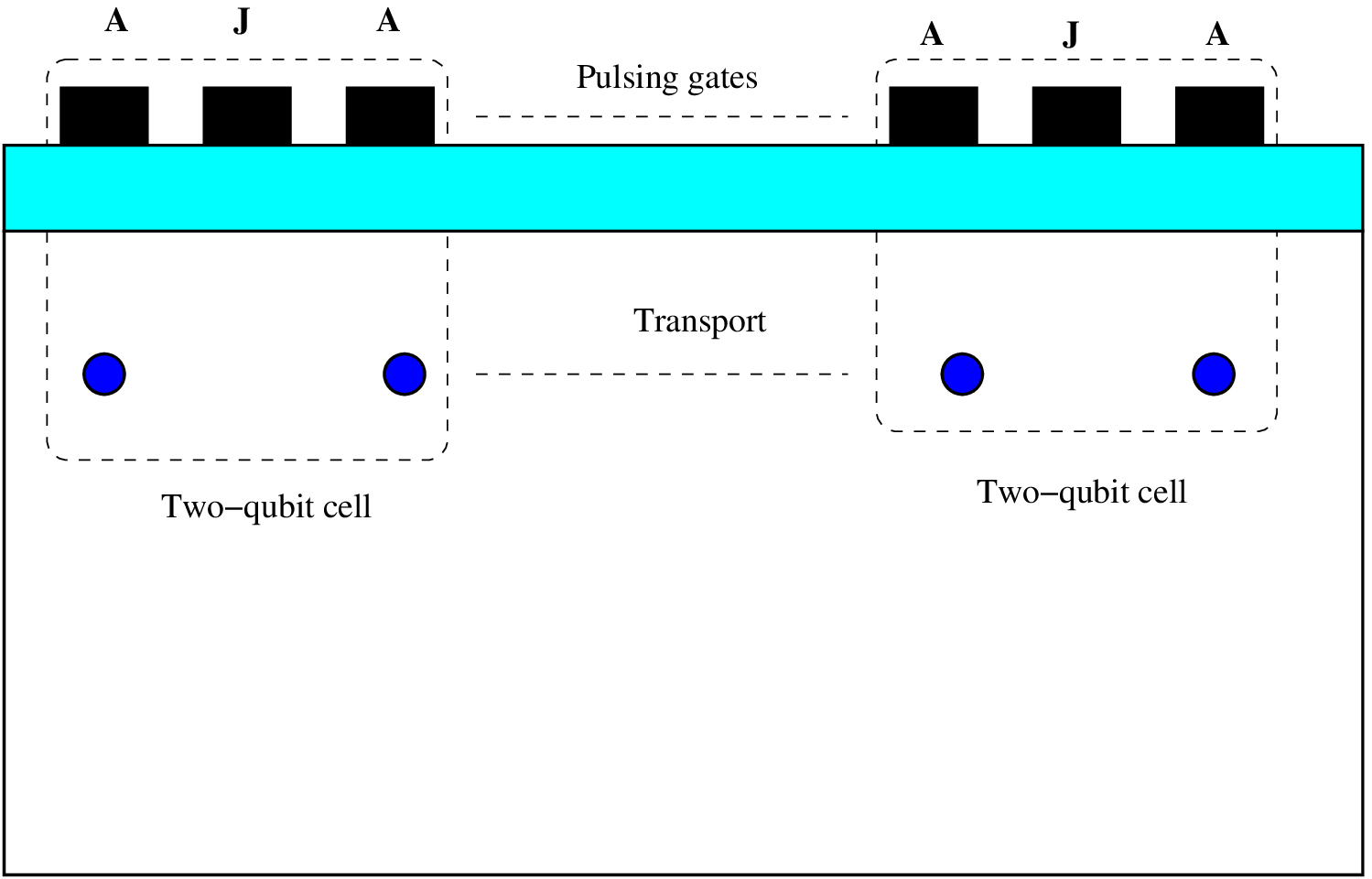}\quad\quad
\includegraphics[height=2in]{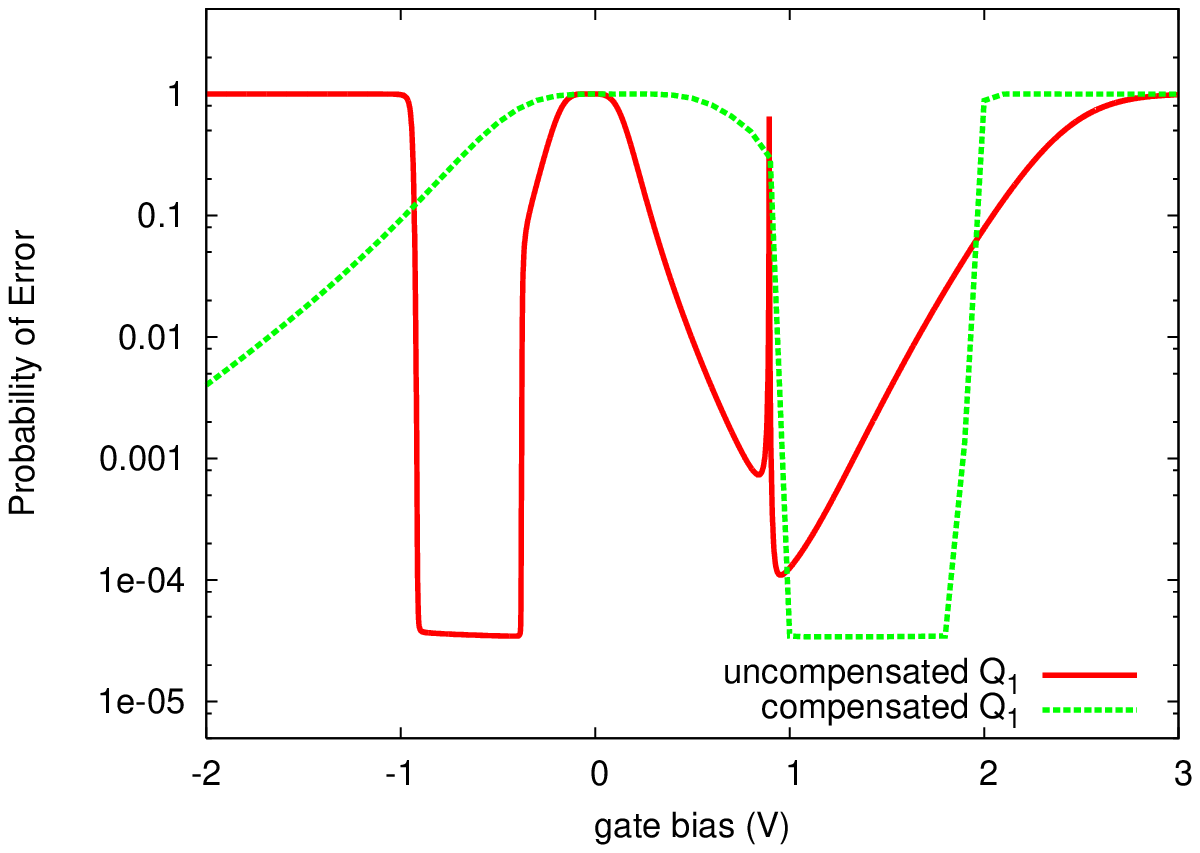}}
\caption{a) Donors organised as two-qubit cells  interacting via quantum transport mechanisms, b) Error probabilities of rotating non-target qubit when the transverse RF is fixed at the resonant frequency of target nucleus. }
\label{2qbit_cell}
\end{figure}

In light of this we consider an architecture which consists of a series of well separated two-qubit cells, where quantum information can be transported between cells via coherent electron transport \cite{Skinner,Greentree}, as sketched in (Fig.~\ref{2qbit_cell} a)). We have calculated the fidelity of single qubit operations for both the uncompensated, and a compensated gate profile, which we show in (Fig.~\ref{2qbit_cell} b)). This gives a more robust compensated voltage regime in which to perform single qubit operations.

\section{Conclusion}

 Practical quantum devices require scalable architectures where crosstalk would have to be minimised. The feasibility of the conventional Kane architecture has been analysed in light of such problems that arise due to close proximity of qubits. The concept of cross-talk in dopant-qubit architectures has been introduced and the physics behind this effect is explained.  We have shown that cross-talk can be reduced using gate voltages based on linear compensation techniques, however, the use of more sophisticated quantum control techniques may be also be beneficial.

In response to the various issues for scale-up of the original Kane proposal, which includes the problem of cross-talk analysed here, an alternative quasi 2D donor based architecture has been proposed \cite{Hollenberg4}. The initial analysis of cross-talk for the two donor interaction cells shows that the problem can be more effectively compensated in such an architecture. 

\section{Acknowledgment}

G.K. would like to thank J. H. Cole and T. R. Starling for useful discussions. This work was supported by the Australian Research Council, the Australian Government and by the US National Security Agency (NSA), Advanced Research and Development Activity (ARDA), and the Army Research Office (ARO) under contract number W911NF-04-1-0290.

\end{document}